\pdfoutput=1
\documentclass[%
prx,twocolumn,
notitlepage,
superscriptaddress,
amsmath,
amssymb,
aps,
prx,
floatfix,
longbibliography
]{revtex4-1}

\usepackage{color}
\usepackage{amsmath}
\usepackage{physics}
\usepackage{amssymb}
\usepackage{bm}
\usepackage{graphicx}
\usepackage[usenames,dvipsnames]{xcolor}
\usepackage{xcolor}
\usepackage[%
  colorlinks=true,
  urlcolor=blue,
  linkcolor=blue,
  citecolor=blue
]{hyperref}
\usepackage{caption}
\usepackage{array}
\definecolor{bleu}{rgb}{0.16,0.2.5,0.36}

\newcommand{\sj}[1]{\textcolor{blue}{#1}}

\begin{document}
\pdfoutput=1


\title{Slow kinks in dissipative kirigami}

%

%
\author{Shahram Janbaz}
\affiliation{Institute of Physics, Universiteit van Amsterdam, 1098 XH Amsterdam, The Netherlands}
\author{Corentin Coulais}
\affiliation{Institute of Physics, Universiteit van Amsterdam, 1098 XH Amsterdam, The Netherlands}

\begin{abstract}
Mechanical waves that travel without inertia are often encountered in nature---e.g. motion of plants---yet such waves remain rare in synthetic materials. Here, we discover the emergence of slow kinks in overdamped metamaterials and we show that they can be used for applications such as sensing, dynamic pattern morphing and transport of objects. To do this, we create dissipative kirigami with suitably patterned viscoelasticity. These kirigami shape-change into different textures depending on how fast they are stretched. We find that if we stretch fast and wait, the viscoelastic kirigami can eventually snap from one texture to another. Crucially, such a snapping instability occurs in a sequence and a travelling overdamped kink emerges. We demonstrate that such kink underpins dynamic shape morphing in 2D kirigami and can be used to transport objects. Our results open avenues for the use of slow kinks in metamaterials, soft robotics and biomimicry.
\end{abstract}

\maketitle
\begin{figure}[b!]
\includegraphics[width=0.48\textwidth]{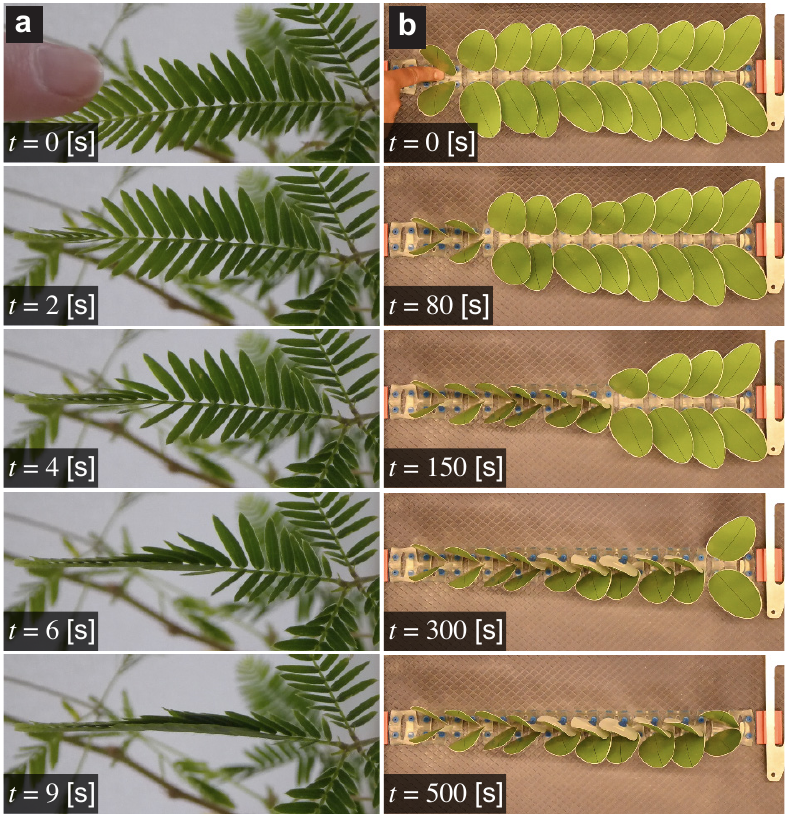}
\caption{{\bf Bio-mimicking slow kinks.} (a) The leaflets of a Mimosa Pudica plant sense their environment by exhibiting a sequential folding process once touched. (b) A viscoelastic kirigami strip carrying paper leaflets shows a similar sequential folding after a touch.}
\label{fig:F1}
\end{figure}

Travelling mechanical kinks are met across a wide range of scales in materials science, from ferro-electrics and shape-memory alloys undergoing phase transitions \cite{zhang2005computational,falk1984ginzburg,kochmann2017exploiting} to flexible metamaterials guiding nonlinear waves \cite{kochmann2017exploiting, deng2021nonlinear, jin2020guided, nadkarni2014dynamics, deng2020pulse, raney2016stable, yasuda2020transition, hwang2021extreme, hwang2021topological}. Yet, in most cases, such waves require inertia to travel. They cannot travel by themselves in overdamped media, unless the material is active~\cite{scheibner2020odd,braverman2021topological, poncet2022soft} or stimuli-responsive~\cite{gu2020magnetic,dong2019multi, chen2018harnessing}. The latter case of stimuli-responsive materials is particularly common in nature, such as in plants that respond to humidity or to touch~\cite{zhang2022unperceivable,reyssat2009hygromorphs,nakamura2013jasmonates,guo2015fast}. A noteworthy example is that of Mimosa Pudica~\cite{bose1914iii, ricca1926transmission, sibaoka1969physiology, hagihara2022calcium} (\textcolor{blue}{Fig.}~\ref{fig:F1}\textcolor{blue}{a} and \sj{Video V1}): when touching one of its leaves, one can observe that its leaflets sequentially fold up in a few seconds, thus giving rise to a travelling mechanical wave. This is a nonlinear wave: the closure and opening of the leaflets can be described as transformation between two stable configurations~\cite{forterre2005venus,dumais2012vegetable,forterre2013slow} and the propagation of sequential folding is understood to be a reaction-diffusion process~\cite{awan2018communication}. Inspired by this phenomenon, we ask here whether one can engineer a similar nonlinear mechanical wave without inertia.

To do this, we consider a kirigami metamaterial~\cite{rafsanjani2019propagation,rafsanjani2017buckling,choi2019programming,yang2018multistable}, that we spatially pattern with a suitable combination of viscoelastic materials, and that we decorate with leaflets (\textcolor{blue}{Fig.}~\ref{fig:F1}\textcolor{blue}{b},
\sj{Video V1}, and see ESI). As with Mimosa Pudica, a slight nudge leads to a sequential folding of the leaflets and a travelling mechanical wave, but this time it takes a few minutes. In this case, it is the combined effect of viscoelastic relaxation and snapping instability that triggers the wave. In both cases of Mimosa Pudica and of dissipative kirigami, inertia is irrelevant and such wave is overdamped. In this article, we lay out the design principles for such slow waves and demonstrate that they can be used to achieve dynamic shape-morphing and transport of objects. 

\begin{figure*}[t!]
\includegraphics[width=0.95\textwidth]{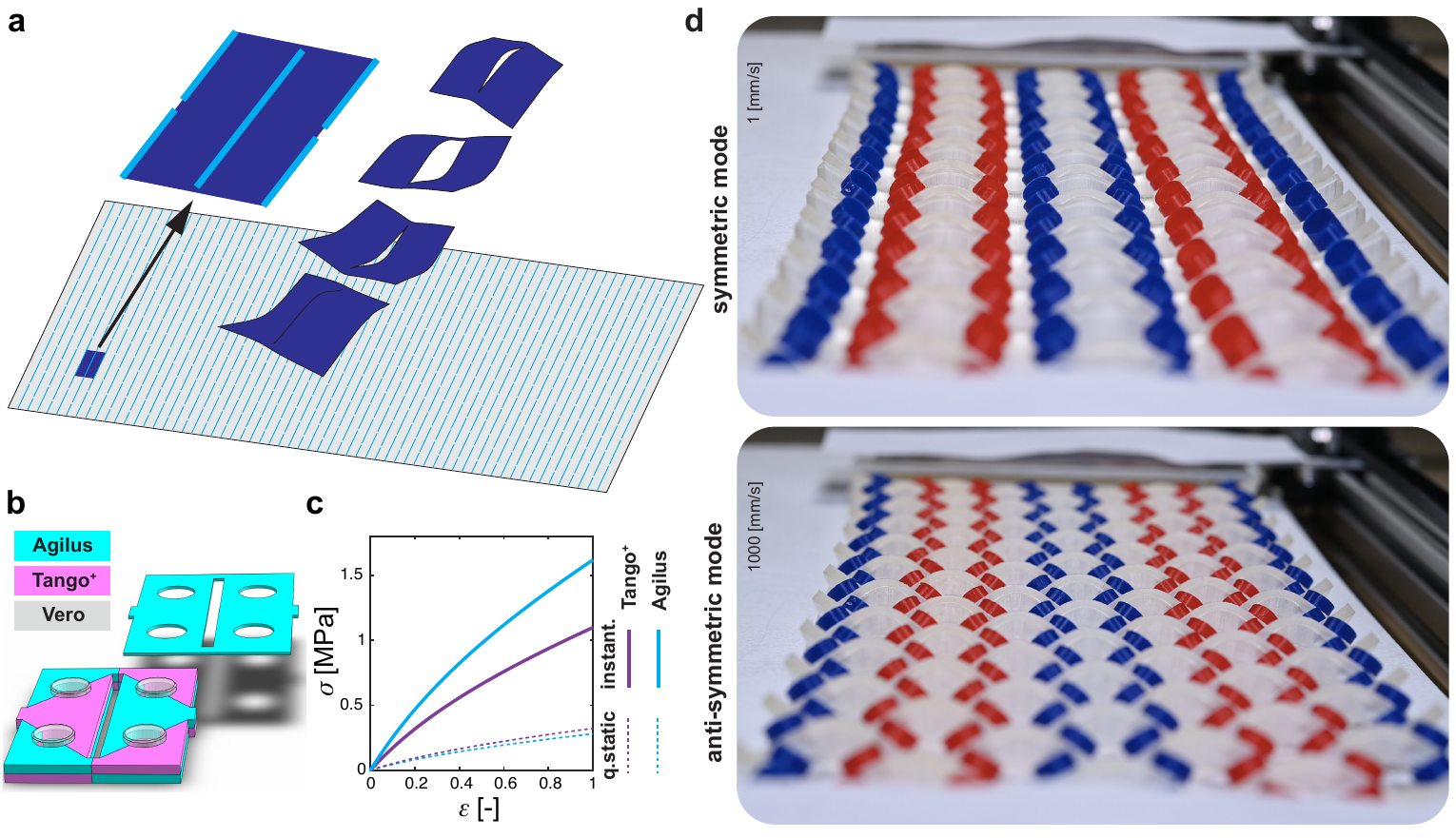}
\caption{{\bf Multi-texture kirigami.} (a) The behavior of kirigami made by introducing regular parallel cut-lines can be described using one of its representative unit cells. Such unit cells may deform in two (symmetric and anti-symmetric) ways upon stretch. (b) We propose a multi-texture design of kirigami unit cells in order to control the direction of buckling. (c) We use photopolymers with a sufficient degree of difference in their viscoelastic properties to print viscoelastic kirigami (see ESI for the benefit of geometrical imperfections). (d) A viscoelastic kirigami made by regular patterning of a {TATA} unit cell (made of Tango|Agilus|Tango|Agilus, see ESI) exhibits shape-transformations depending on the applied loading rate}
\label{fig:F2}
\end{figure*}
Our approach uses a kirigami that intrinsically allows for two buckling modes(\textcolor{blue}{Fig.}~\ref{fig:F2}\textcolor{blue}{a}), hence two textures  \cite{yang2018multistable}. In contrast with existing kirigami, which consists of a single material, our kirigami consists of two materials. The two materials are strategically patterned within the unit cell and crucially through its thickness (\textcolor{blue}{Fig.}~\ref{fig:F2}\textcolor{blue}{b}). We use two rubbery materials with distinct viscoelastic relaxation strengths to tune the imperfections that control the direction of buckling in response to the applied loading rate~\cite{bossart2021oligomodal,janbaz2020strain, stern2018shaping,Janbaz_3D_Printable,dykstra2022extreme} (See ESI).
In contrast with single-material kirigami, whose buckling mode is hard to control~\cite{yang2018multistable,dias2017kirigami}, the buckling mode of our multimaterial viscoelastic kirigami is tunable on-the-fly via the rate at which the kirigami is
stretched: at low loading rates, the kirigami buckles into a symmetric mode whereas at high loading rates, it buckles into an anti-symmetric mode, \textcolor{blue}{Fig.}~\ref{fig:F2}\textcolor{blue}{c} and \sj{Video V2}).
\begin{figure*}[]
\centering
{\includegraphics[width=0.95\textwidth]{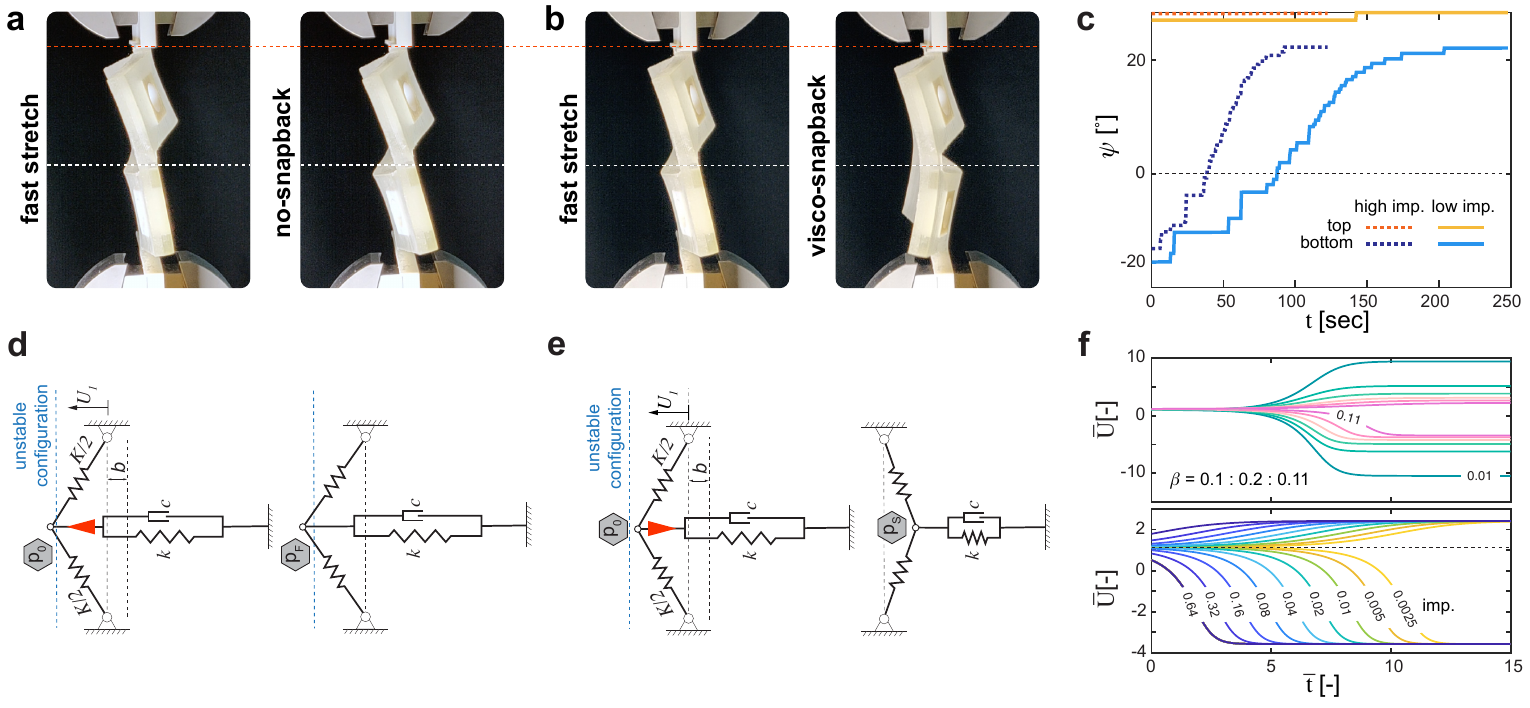}}
\caption{{\bf Viscoelastic snapping.} Upon a fast stretch, the geometry of a TATA unit cell deforms (buckle) anti-symmetrically (see ESI). (a,b) Within certain ranges of stretch, the strip(s) of the buckled unit cell exhibits a viscoelastic snap-back to its lower speed mode---symmetric mode. (c) We used a 3D tracking technique (see ESI) to quantify the role of geometrical imperfections on the acceleration of snap-back. (a,b) We use a viscoelastic von Mises truss configuration to represent the condition for the snap-back of a kirigami uni cell from an unstable high-speed to a lower-speed mode (see ESI). (d) Numerically solving the equation of motion (\ref{eq:viscosnap}), regardless of the effect of the initial condition (i.e., geometrical imperfection), shows that the amplitude of snap-back is bonded to the system parameter \(\beta\). On the other side, the level of imperfection highly influences the time period prior to a viscoelastic snap-back}

\label{fig:F3}
\end{figure*}

Such kirigami exhibits viscoelastic stress relaxation and for some regimes of strain a viscoelastic snap-back (Fig. ~\ref{fig:F3}\textcolor{blue}{a,b} and \sj{Video V3}) is anticipated \cite{gomez2019dynamics, santer2010self,dykstra2022extreme}. We quantified such snapping by tracking the relaxation of a unit cell stretched at a high speed in 3D (see ESI). While at short times, the kirigami unit cell is deformed into the high-speed anti-symmetric mode, at longer times, it gradually deforms into a symmetric mode. Such transformation starts with a slow creep process, but over the course of $100$~sec, the bottom half of the unit cell snaps back, such that the unit cell ultimately relaxes into the low-speed symmetric mode (Fig.~\ref{fig:F3}\textcolor{blue}{c}, blue curve).
In a limited range, the snap-back of viscoelastic strips is highly sensitive to geometrical imperfections (see ESI). A slight change in the angle of rigid end-connections of the unit cell result in, for example, a shorter delay prior to a viscoelastic snap-back while there is not a visible change in the final angle \(\psi\) between the panels (\textcolor{blue}{Fig.}~\ref{fig:F3}\textcolor{blue}{c}, dashed curve). That shows the strong influence of geometrical imperfections~\cite{gomez2019dynamics, janbaz2020strain} on the viscoelastic snap-trough of our kirigami.

To capture these observations and quantify the effect of material, geometry, and imperfections on viscoelastic snapping, we use a viscoelastic von Mises truss (\textcolor{blue}{Fig.}~\ref{fig:F3}\textcolor{blue}{d,e}) to model switching between symmetric and anti-symmetric buckling modes. The truss is made by jointing a pair of linear springs \(K/2\) and a pair of parallel spring \(k\) and dashpot \(c\) while all the spring are pre-strained. Using a Lagrangian formulation that comprises the elastic potential energy \(V\) and the Rayleigh dissipation function \(D\)~\cite{ginsberg1998advanced}, we obtain the non-dimensional equation of motion (see ESI):
\begin{equation}
\frac{\partial\bar{U_1}}{\partial \bar{t}} = \bar{U}_{1} - \beta \bar{U}_{1}^3 - 1 ,
\label{eq:viscosnap}
\end{equation}
\noindent 
in which $\bar{U}_{1}$ and $\bar{t}$ are the dimensionless displacement and time. The dimensionless parameter $\beta = \frac{Kk^2b^2d_0} {2((-K-k)a+Kd_0)^3}$ is a function of material properties and nonlinear geometry, where \(d_0\) is the natural length of the springs \(K/2\), \(a\) is half of the distance between the fixed joints, and \(b\) is the offset point \(p\) from and straight configuration while the spring \(k\) is relaxed. We can then readily use Eq.~(\ref{eq:viscosnap}) to obtain the condition for viscoelastic snap-back: ($0 < \beta < \frac{4} {27}$) \cite{kurosh1980higher}, which is the condition for bi-stability. The parameter $\beta$ not only determines whether the unit cells snaps, it also determines the magnitude of snap-back (\textcolor{blue}{Fig.}~\ref{fig:F3}\textcolor{blue}{f}): the smaller $\beta$, the larger the magnitude of snapping. In contrast, the level of imperfection determines the time it takes for snapping to occur (\textcolor{blue}{Fig.}~\ref{fig:F3}\textcolor{blue}{f}): the more the imperfection, the shorter the initial creep and the quicker the snapping. Qualitatively, the viscoelastic snapping of a unit cell can be described as follow: at short times, the unit cell is metastable, and might be trapped in the up-state, which slowly creeps. However, at longer times, the unit cell becomes monostable: the up-stable is not stable anymore, hence the unit cell snaps into the down-state.


\begin{figure*}[t!]
\centering
{\includegraphics[width=0.95\textwidth]{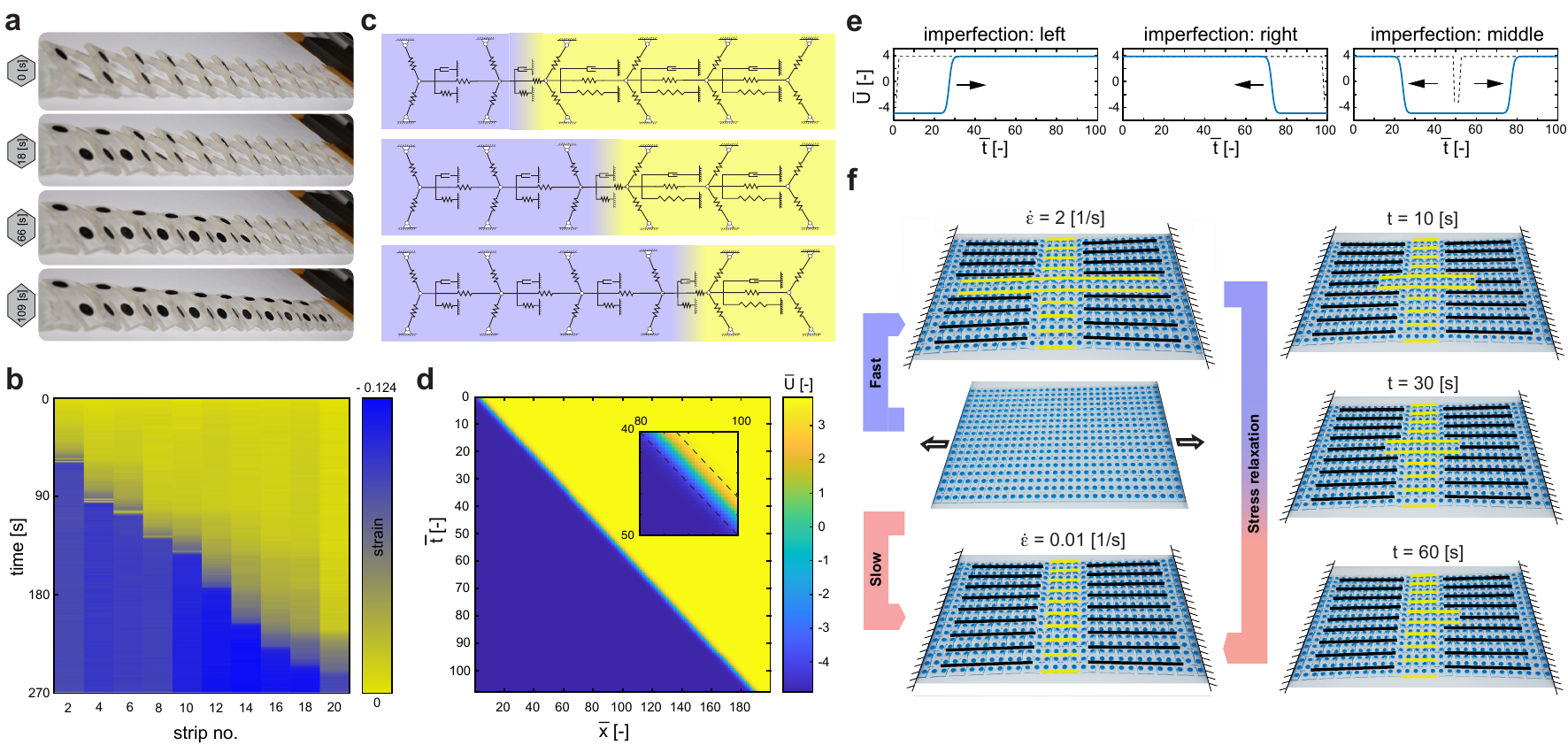}}
\caption{{\bf Slow kink in pre-stretched kirigami strands.} (a) A kirigami strand consisting of twelve identical TATA unit cells exhibits a travelling kink (see ESI and \sj{Video V4}). (b) Measuring the values of lateral strains (see ESI), corresponding to the snapping of the even index strips reveals the emergence of a constant speed wave. (c) We use a series of viscoelastic trusses integrated using linear springs \(R\) to represent a kirigami strip. (d) Numerically solving the reaction-diffusion equation~(\ref{eq:viscosnap1D}) for \(\beta = 0.05\) reveals the emergence of a constant speed wave. The high- and low-speed configurations of the individual viscoelastic trusses are then corresponding to the maximum and minimum roots of the reaction function \(f(\bar{U})=\bar{U}_{1} - \beta \bar{U}_{1}^3 - 1\) respectively (i.e., \(\bar{U} \approx +3.843\) and \(-4.907\)). (e) The direction of travelling kinks depend on the initial imperfections. (f) A strain-rate dependent 2D shape-morphable plate exhibit dynamic shape transformation from its high-speed to low-speed mode}
\label{fig:F4}
\end{figure*}


We then hypothesize that similar to the propagation of inertial kinks in bi-stable chains  \cite{nadkarni2014dynamics}, our viscoelastic trusses in series will also exhibit kinks, but we expect the dynamics of these kinks to be overdamped.
To examine this assumption, we build a kirigami strip, prestretch it fast such that it buckles into the anti-symmetric mode and let it relax (see ESI). Our experiments confirm the emergence of a localized mechanical kink starting from the high-speed anti-symmetric mode, into the low-speed symmetric mode (\textcolor{blue}{Fig.}~\ref{fig:F4}\textcolor{blue}{a} and \sj{Video V4}). The mechanical kink travels from one end to the another end side of the kirigami. 

To quantify the dynamics of such slow kink, we measured the instantaneous lateral strain of the half unit cell (See ESI, \sj{Video V5}), which undergoes a jump upon snapping (\textcolor{blue}{Fig.}~\ref{fig:F4}\textcolor{blue}{b}). The sequence of such snapping events reveals the existence of a kink, travelling at constant velocity (around 1 $mm/s$).
Clearly, the timescale is much higher than the speed of vibrations and the slow kink undergoes purely dissipative dynamics. This dynamics can be understood as the propagation of an imperfection. 


To describe this mechanism, we promote our model of a single unit cell Eq.~(\ref{eq:viscosnap}) to a one-dimensional strip. We model a viscoelastic kirigami strip as a series of lumped viscoelastic trusses, connected by linear springs \(R\) (\textcolor{blue}{Fig.}~\ref{fig:F4}\textcolor{blue}{c}). The dimensionless equation of motion, excluding the effect of inertial forces, in continuum limit can be then expressed as (see ESI):
\begin{equation}
\frac{\partial\bar{U}}{\partial \bar{t}} = \bar{U} - \beta {\bar{U}}^3 -1 + \pdv[2]{\bar{U}}{\bar{X}} 
\label{eq:viscosnap1D}
\end{equation}
in which $\bar{X}$ is the nondimensionalized coordinate. Eq.~(\ref{eq:viscosnap1D}) is a reaction-diffusion equation that is normally used to describe transitions in bi-stable media~\cite{mcdougal2013reaction} and are well known to host kinks~\cite{deng2021nonlinear,jin2020guided}, but that describes in our case the mechanical analogue of such reaction-diffusion kinks.
Numerical solutions of Eq.~(\ref{eq:viscosnap1D}) confirm the emergence a kink travelling at a constant velocity, that is a moving transition between the initial metastable up-state ($\bar{U}_{+}$) and the stable down-state ($\bar{U}_{-}$) (\textcolor{blue}{Fig.}~\ref{fig:F4}\textcolor{blue}{d,e}). Independent from the limits that may arise from friction and anisotropy in our experiments, our model represent a sensitive system that exhibit travelling kinks depend on how it is perturbed. Therefore, theoretically, the direction of front wave can be specified based on where it has been touched (\textcolor{blue}{Fig.}~\ref{fig:F4}\textcolor{blue}{e} and \sj{Video V6}).

The combination of experiments with this model provides the following physical picture:
the time delay before the first snapping event (i.e., corresponding to strip no. 2) can be attributed to the size of imperfection at the boundary while the rest of unit cells are still slowly creeping in their high-speed anti-symmetric mode (\textcolor{blue}{Fig.}~\ref{fig:F4}\textcolor{blue}{b}). When the first unit cell snaps, it creates a larger imperfection for its neighbor and accelerates its snapping. This snapping of this second unit cell will in turn create a larger imperfection and accelerate the snapping of the third unit cell and so on. This sequential mechanism ultimately creates a travelling kink with purely overdamped dynamics.
Now that we have captured the essence of this dissipative kink, one can apply it to obtain shape-changing functionalities \cite{celli2018shape,an2020programmable}. We, therefore, demonstrate that viscoelastic snapping can be used to obtain dynamic shape morphing on 2D kirigami (\textcolor{blue}{Fig.}~\ref{fig:F4}\textcolor{blue}{f}, \sj{Video V7}, and see ESI).
To do this, we modulate the strain-rate dependency and the snapping properties of kirigami unit cells by slightly varying their geometry and by rationally patterning the viscoelastic material in the kirigami (ESI), such that the high-speed mode of the kirigami is in the shape of a ``plus'' sign and the low-speed more in the shape of a ``minus'' sign. 
We then confirm experimentally that this kirigami exhibits a dynamic shape-transformation, from a plus- to a minus-textured pattern, subsequent to a fast stretch (\textcolor{blue}{Fig.}~\ref{fig:F4}\textcolor{blue}{f}, \sj{Video V7}). In our experiments, we observe a progressive shape transformation, starting from the two clamped sites of the kirigami plate, that results in merging of the textures at the two side of the minus strip. This transformation is mediated by a slow dissipative kink described above.
Finally, the kirigami adopts into a minus sign texture which is the same as what we observed in the slow-speed experiments. 

\begin{figure}[t!]
\centering
\includegraphics[width=0.47\textwidth]{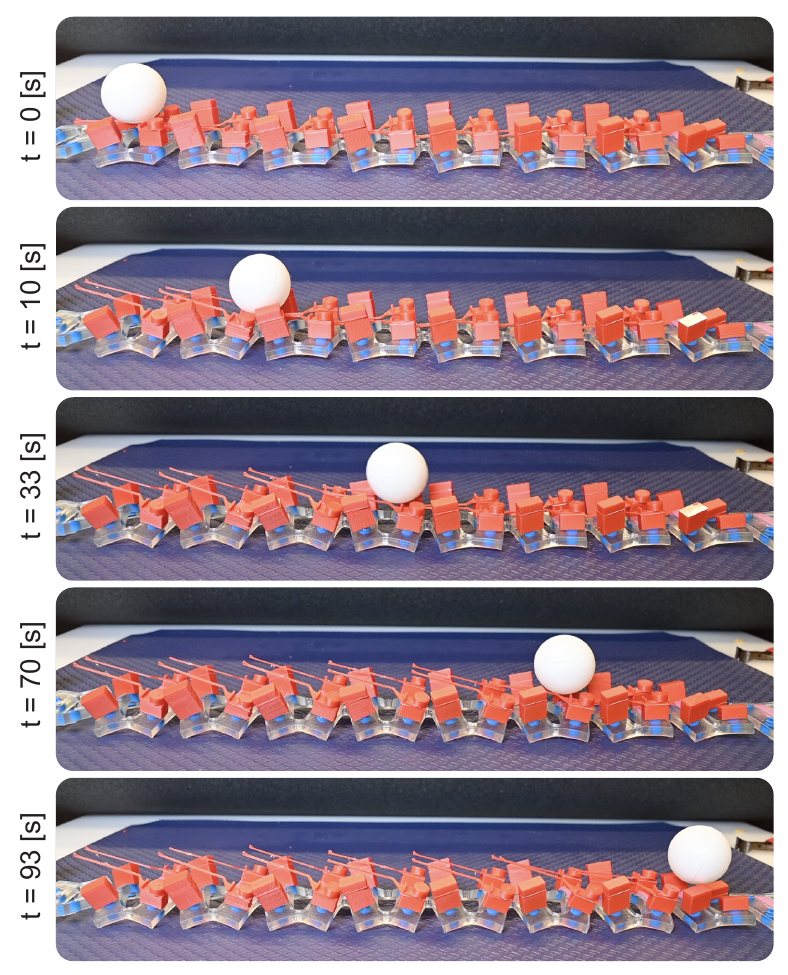}
\caption{{\bf Slow kinks move objects.} A kirigami strip equipped with arm-links and guiders carry a ping-pong ball while initially stretch at a high speed}

\justifying
\label{fig:F5}
\end{figure}

Besides the change in the geometry of kirigami, we demonstrate that this dissipative kink can also be used to perform a basic mechanical task, for example, to transport an object forward~\cite{demirors2021amphibious}. For this purpose, we dress a kirigami strip with arm-links and guiders (\textcolor{blue}{Fig.}~\ref{fig:F5}, \sj{Video V8}, and see ESI). As described above, such kirigami, after being stretched at a high speed and set to relax, exhibits a travelling kink. This kink lifts the arm-links and guiders sequentially and carry an object---in this case a ping-pong ball---forward.

In summary, we have shown that viscoelasticity in kirigami structures can be harnessed to create bistable mechanisms, whose snapping is mediated by creep and that in turn exhibit overdamped travelling kinks. These slow kinks can in turn be used to create complex 2D shape morphing systems and mechanisms with the ability of mass transport. In contrast with inertial kinks \cite{deng2020pulse, jin2020guided, kochmann2017exploiting, deng2021nonlinear,nadkarni2014dynamics}, overdamped kinks can propagate arbitrarily slowly. This feature facilitates the design of materials that mimic the motion of plants~\cite{mcdougal2013reaction}, and hence adds a concept to the toolbox of soft robots \cite{rus2015design,chen2018harnessing}.

\bigbreak
\emph{\bf{Acknowledgments.}} 
We thank Freek van Gorp, Jonas Veenstra, Keivan Narooei and Mazi Jalaal for insightful discussions, Anastasiia Krushynska for real-time imaging of Mimosa Pudica, Clint Ederveen Janssen, Daan Giesen, Kasper van Nieuwland, and Ronald Kortekaas for their technical support. We acknowledge funding from the Netherlands Organisation for Scientific Research under grant agreement NWO TTW 17883 and from the European Research Council under grant agreement 852587.

\bibliographystyle{naturemag}
\bibliography{Reference}

\clearpage
\setcounter{equation}{0}
\renewcommand{\theequation}{A\arabic{equation}}
\setcounter{figure}{0}
\renewcommand{\thefigure}{S\arabic{figure}}


\begin{widetext}

\subsection*{Electronic supplementary information to:}

\subsection*{\huge Slow kinks in dissipative kirigami}

\bigbreak
\subsection{Geometrical design of kirigami plates}\label{A}
In our study, we use a thick kirigami that allows multimaterial texturing through its thickness in order to achieve strain rate dependency. The simple geometry of our kirigami design---a plate perforated with a regular pattern of parallel cut-lines---can be represented using one of its unit cells (\textcolor{blue}{Figs.}~\ref{fig:F2}\textcolor{blue}{a},~\ref{fig:S1}\textcolor{blue}{a}). We split the geometry of the representative unit cell into four main compartments for multimaterial texturing (\textcolor{blue}{Figs.}~\ref{fig:F2}\textcolor{blue}{b},~\ref{fig:S1}\textcolor{blue}{b}). In general, upon stretch, such a unit cell randomly buckles into two distinct modes: symmetric and anti-symmetric modes (\textcolor{blue}{Fig.}~\ref{fig:F2}\textcolor{blue}{a}). Patterning with multimaterial allows us then to overcome this randomness and to control the direction of buckling by exploiting imperfections that arise from the behavior of the chosen viscoelastic polymers~\cite{bossart2021oligomodal,janbaz2020strain,dykstra2022extreme}. We computationally examined and analyzed the geometrical features of the representative unit cell used in our study, to ensure the manufacturability of strain rate sensitive kirigami by considering the limits of the commercially available photopolymers~\cite{Janbaz_3D_Printable} (see section \ref{B}). We were, therefore, able to manufacture viscoelastic kirigami plates that transform their geometry in different fashions in response to predicted ranges of strain rates using a polyjet 3D printer (Connex 500, Stratasys) with two soft viscoelastic photopolymers (Tango\(^{+}\) and Agilus, Stratasys). Given the fact that these two polymers exhibit different levels of shrinkage after photo-polymerization~\cite{zhao2017origami}, the final 3D printed material has a wavy surface with maximum curvatures close to the symmetry lines of its unit cells. In the absence of extra photo-curable soft polymers, the resulting geometrical imperfections are useful to prevent snapping in half unit cells with top Agilus material once the paired half unit cell with top Tango\(^{+}\) exhibits viscoelastic snapping. To minimize the size of imperfection, depending on the scale of kirigami unit cells, we use a finishing thin layer of Agilus coating on the multi-texture kirigami. Finally, to further maximize the strain rate dependency in kirigami designs, we replace the area between the symmetry lines of kirigami unit cells with rigid stiff material (Vero, Stratasys) disks. The rigid areas can be also useful for carrying extra mechanisms we use in (\textcolor{blue}{Figs.}~\ref{fig:F1},~\ref{fig:F5}).
\begin{figure}[h]
\centering
\includegraphics[width=0.96\textwidth]{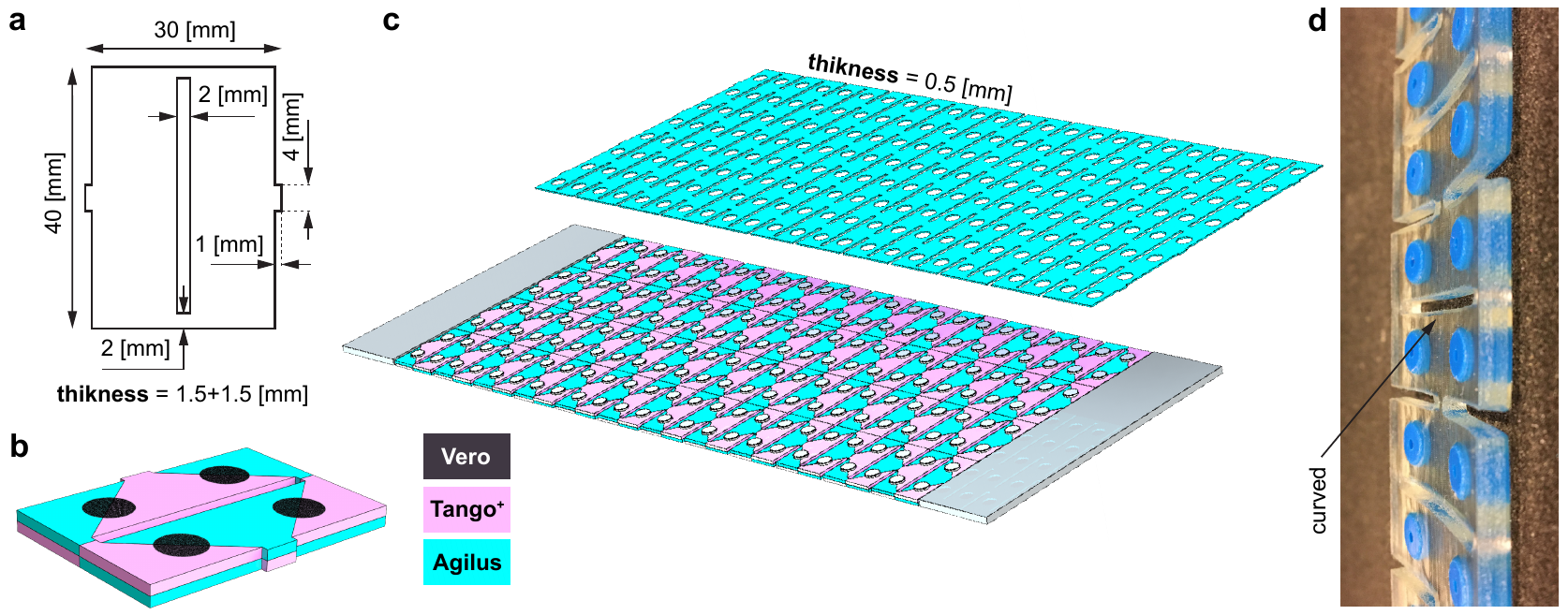}
\caption{{\bf Viscoelastic kirigami plates.} (a) The schematic of the standard unit cell used in our study. (b) We split the geometry of the kirigami unit cell into four equally thick compartments for anisotropic multimaterial texturing---using two soft photopolymers Tango\(^{+}\) and Agilus. In some designs we replace the boundary of Tango\(^{+}\) and Agilus with Vero, which is a stiff photopolymer, to enhance the strain rate dependency and facilitate carrying objects. (c) In the specimens made of standard-size unit cells, we use a thin layer of Agilus coating to minimize the size of geometrical imperfections. (d) The curved surface of a 3D printed kirigami strip at its bottom side}
\label{fig:S1}
\end{figure}
\subsection{Strain rate dependent kirigami plates}\label{B}
In order to predict the manufacturability of viscoelastic kirigami using our conventional additive manufacturing technique (i.e., polyjet printing), we first used non-linear computational mechanics (Abaqus ver. 2020, Standard solver). We analyzed the post-buckling of the units cells we aim to use in our study (\textcolor{blue}{Figs.}~\ref{fig:S2}\textcolor{blue}{a-b}). The commercial materials that we used for the fabrication of kirigami materials are different viscoelastic elastomers (i.e., Agilus and Tango$^{+}$, Stratasys). We, therefore, used a visco-hyperelastic material model (using the first term of Prony series) to define the visco-hyperelasticity of Agilus and Tango$^{+}$. The material constants $C_{10A}^{\infty} = 0.080 $ [MPa] and $C_{10T}^{\infty} = 0.092 $ [MPa], as well as the dimensionless coefficients of Prony series $g_{1A} = 0.826 $ and $g_{1T} = 0.705 $, and time scales \sj{$\tau_{1A} = 0.25 $} and $\tau_{1T} = 0.218 $ have been chosen according to the fitting of the visco-hyperelastic material model to the stress relaxation test results of Agilus and Tango$^{+}$ (\textcolor{blue}{Figs.}~\ref{fig:S2}\textcolor{blue}{d, e}). The material parameters were determined by minimizing the difference between the stress values predicted by the material model and the experimental data, assuming that the two viscoelastic polymers are incompressible \cite{eghbali2022hyperelastic}. Moreover, to discretize the geometry of the kirigami, we used three-dimensional elements C3D8H. A mesh convergence study has been performed to ensure the insensitivity of our computational analysis to the mesh size. The clamped-clamped condition has been realized by fixing the nodal points of one end of the kirigami while the nodal points of the moving end have the freedom to move along the axial direction of the plates. Periodic boundary condition has been satisfied by constraining the motion of the nodal points at the two free boundaries of the unit cells with respect to a reference point placed on one node. Our computational analysis, then, confirm the manufacturability of our multi-texture designs using polyjet printing.

\begin{figure}[h]
\centering
\includegraphics[width=0.96\textwidth]{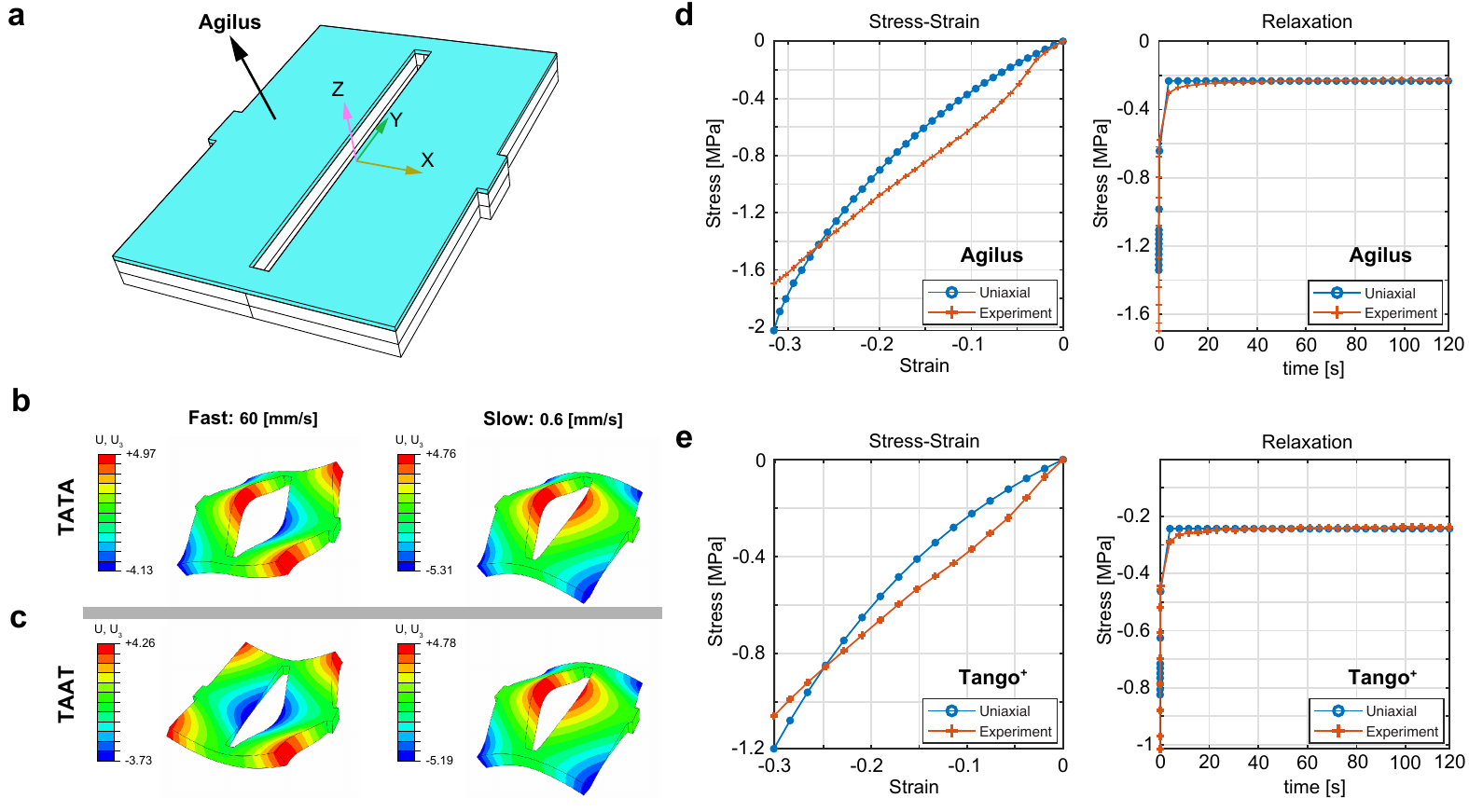}
\caption{{\bf Computational prediction of strain rate dependent post-buckling of TATA and TAAT unit cells.} (a) In our computational analysis we used a standard-size unit cell covered with a finishing Agilus layer. (b, c) Our computational results confirm the manufacturability of TATA and TAAT designs using polyjet printing. (d, e) The fitting of the instantaneous and relaxation data to a single-term Prony series}
\label{fig:S2}
\end{figure}

\subsection{Strain rate sensitivity}
Using our custom-made test bench (\textcolor{blue}{Fig.}~\ref{fig:S3}\sj{a}) we show that our two kirigami designs (TATA and TAAT) exhibit robust shape-transformations at the lower and upper speed limits of our test bench (i.e., \(v_{min}=1 \: mm/s\) and \(v_{max}=1000 \: mm/s\)). From low to fast speeds, the TAAT design exhibits two symmetric modes, and the TATA design exhibits switching from a symmetric to an anti-symmetric geometry (\textcolor{blue}{Fig.}~\ref{fig:F2},~\ref{fig:S3}, and \sj{Video V3, V6}) which are in accordance with our computational predictions. This is expected since we have considered the existing geometrical imperfections and the effect of the cover layer on the dissimilar response of the Agilus and Tango$^{+}$ materials printed at the top and bottom sides of the kirigami unit cells.

At the intermediate range of speeds, we observe that both symmetric and anti-symmetric modes coexist (\textcolor{blue}{Fig.}~\ref{fig:S3}\sj{b}). This coexistence is mainly due to geometrical imperfections and, less importantly, friction and boundary effects. While the TATA design exhibits an irregular mixture of symmetric and anti-symmetric modes, the TAAT design displays a longitudinally aligned bi-domain (\textcolor{blue}{Fig.}~\ref{fig:S3}\sj{b}, \sj{Video V7}). The existence of the bi-domain is mainly due to symmetry and the collective longitudinal behavior of the unit cells. To analyze the sensitivity of buckling to loading speed, using a Matlab code, we evaluated the average values of the lateral strains by detecting and tracking the dark particles placed symmetrically over the geometry of the unit cells (\sj{Video V8}). Independent from the effect of material properties, both designs exhibit an identical pattern of buckling (i.e., symmetric buckling) with similar value of lateral strains ($\varepsilon = -30\%$) when they are stretched at the low speed. At intermediate speeds, \(1<log( \frac{v}{v_{min}})<2\), the buckling of both designs is sensitive to loading speed. The sensitivity can be manifested by a reduction in the absolute value of lateral strains. It can be seen that the TATA design exhibits sensitive buckling over a wider range of speeds (\textcolor{blue}{Fig.}~\ref{fig:S3}\sj{c}) that is mainly due to its anti-symmetric multimaterial pattern.
Eventually, the high-speed mode of buckling appears at higher speeds. While the average value of the lateral strain growth to a positive value ($\varepsilon = 10\%$) in symmetric designs, the maximum value of strain is equal to the average of the lower and higher speeds strains ($\varepsilon = -10\%$). It is noteworthy to note that the analysis of lateral strains is influenced by the fact that the traced particles are placed on top of the kirigami plates. This can be a valuable point while a kirigami plate serves as a flexible substrate for carrying 3D objects as the source of mechanical functionalities.

\begin{figure}[h]
\centering
\includegraphics[width=0.96\textwidth]{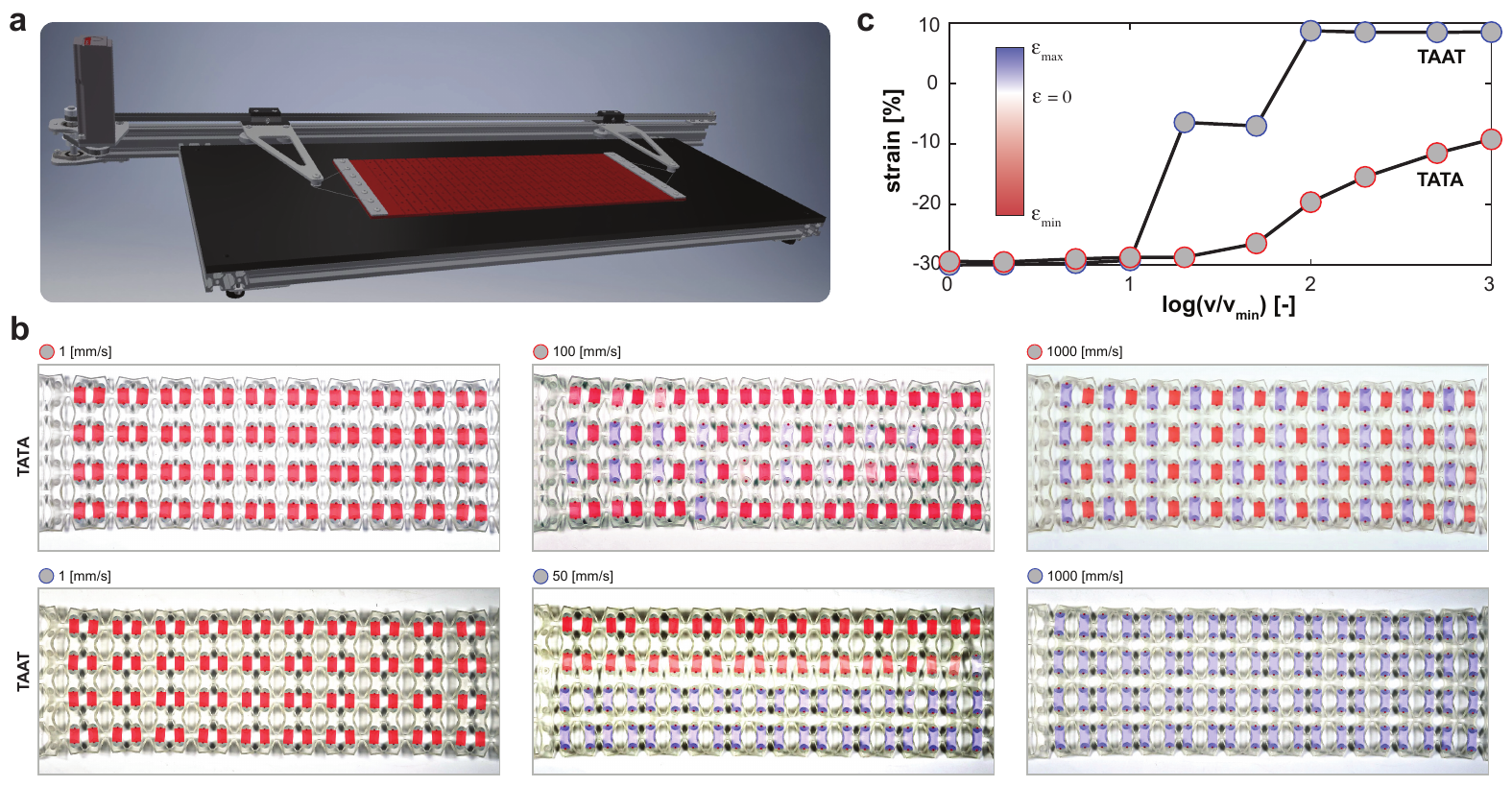}
\caption{{\bf Strain rate sensitivity.} (a) The 3D configuration of fast-stretch setup equipped with a Teflon substrate. (b) The buckled geometry of TATA and TAAT kirigami stretched at different speeds. At intermediate range of speeds combined modes such as bidomain mode is observed. (c) Strain rate sensitivity of TAAT and TATA plates can be manifested according to the intermediate values of lateral strain between the maximum and minimum values of lateral strain}
\label{fig:S3}
\end{figure}

\begin{figure}[h]
\centering
\includegraphics[width=0.98\textwidth]{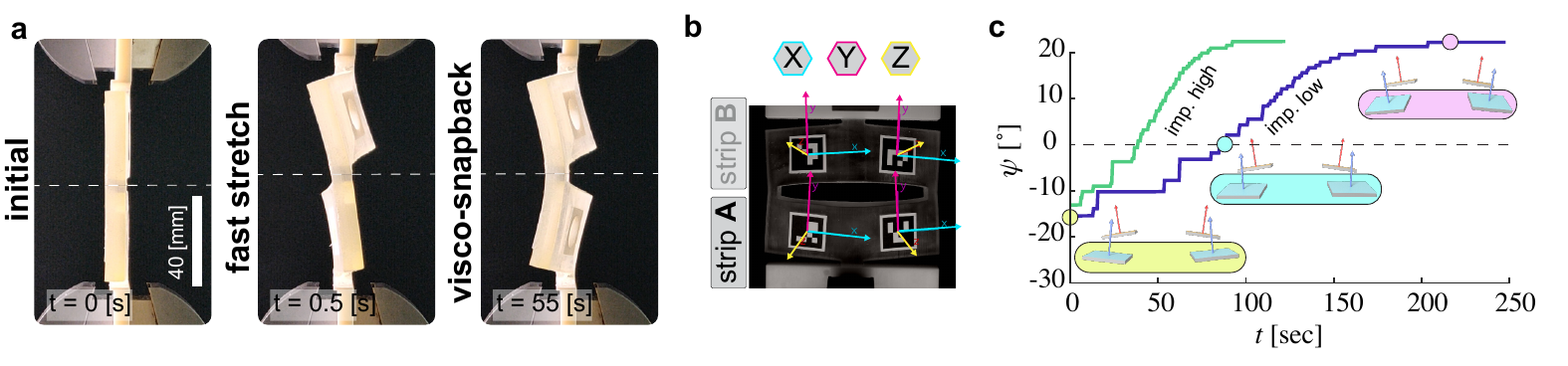}
\caption{{\bf Viscoelastic snap-back.} (a) A TATA unit cell (300\% of standard size, stretched 12.5 mm at 500 mm/s) exhibits a viscoelastic snap-back. (b) We used a tracking system to quantify the viscoelastic behavior of the TATA unit cell. (c) The overdamped snap-back of Strip'A' of the TATA unit cell (300\% of standard size, stretched 12.5 mm at 500 mm/s) is quantified by the change of the angle between the normal vectors on the corresponding markers from positive to negative}
\label{fig:S4}
\end{figure}

\subsection{Viscoelastic snap-back leading to traveling waves}
Viscoelastic kirigami exhibit stress relaxation and for some regimes of strain, a viscoelastic snap-back (Fig. ~\ref{fig:S4}\textcolor{blue}{a} and \sj{Video V3}) is anticipated \cite{gomez2019dynamics, santer2010self,dykstra2022extreme}. We quantify such snapping by tracking the relaxation of a TATA unit cell (300\% of standard size unit cells, stretched 9 mm at 50 mm/s) using an optical tracking system called AruCo \cite{romero2018speeded,garrido2016generation}.
In our experiments, four rigid tiles are embedded symmetrically within the geometry of the unit cell and marked with so called AruCo markers, which are small square monochrome images with specific patterns (Fig. ~\ref{fig:F3}\textcolor{blue}{b}). The patterns are then used to identify the object and determine both the position and rotation of the markers. We track each tile separately by choosing unique markers on each tile.

While initially, the kirigami unit cell is deformed into the high-speed anti-symmetric mode, at longer times, it gradually creeps back into a symmetric mode (Fig. ~\ref{fig:S4}\textcolor{blue}{a} and \sj{Video V3}). Such transformation starts with a slow creep process, but over the course of $100$~sec, the bottom half of the unit cell snaps back, such that the unit cell ultimately relaxes into the low-speed symmetric mode (Fig. ~\ref{fig:F3}\textcolor{blue}{c}, blue curve). Such transformation can be manifested by switching the sign of the angle between the normal vectors at the middle of corresponding AruCo markers on the snappy half unit cell. 
In a limited range, the snap-back of viscoelastic strips is highly sensitive to geometrical imperfections. A slight change in the angle of rigid end-connections of the unit cell result in, for example, a shorter delay prior to a viscoelastic snap-back while there is not a visible change in the final angle \(\psi\) between the panels (\textcolor{blue}{Fig.}~\ref{fig:S4}\textcolor{blue}{c}, green curve). That shows the strong influence of geometrical imperfections~\cite{gomez2019dynamics, janbaz2020strain} on the viscoelastic snap-trough of our kirigami.

\subsection{Quantifying the wave propagation in kirigami strips}
To quantify the travelling of overdamped waves in our experiments, we measured the lateral straining of the snappy half unit cells, from their high-speed stretched configuration, by tracking two markers on their top sides using a custom Matlab code. While lateral strains exceed more than $6\%$, the snap-back of viscoelastic strips can be manifested based on a rapid change in the value of the lateral strain over time$-$as it is distinguishable according to the color codes in \textcolor{blue}{Fig.}~\ref{fig:S5}\textcolor{blue}{}. The sequence of snap-backs reveals the existence of a mechanical kink wave travelling at a constant speed (\textcolor{blue}{Fig.}~\ref{fig:S5}\textcolor{blue}{b}). Moreover, the time delay prior to the first snap-back (i.e., corresponding to strip no. 2) can be attributed to the size of imperfections at the corresponding boundary. 

\begin{figure}[h]
\centering
\includegraphics[width=0.96\textwidth]{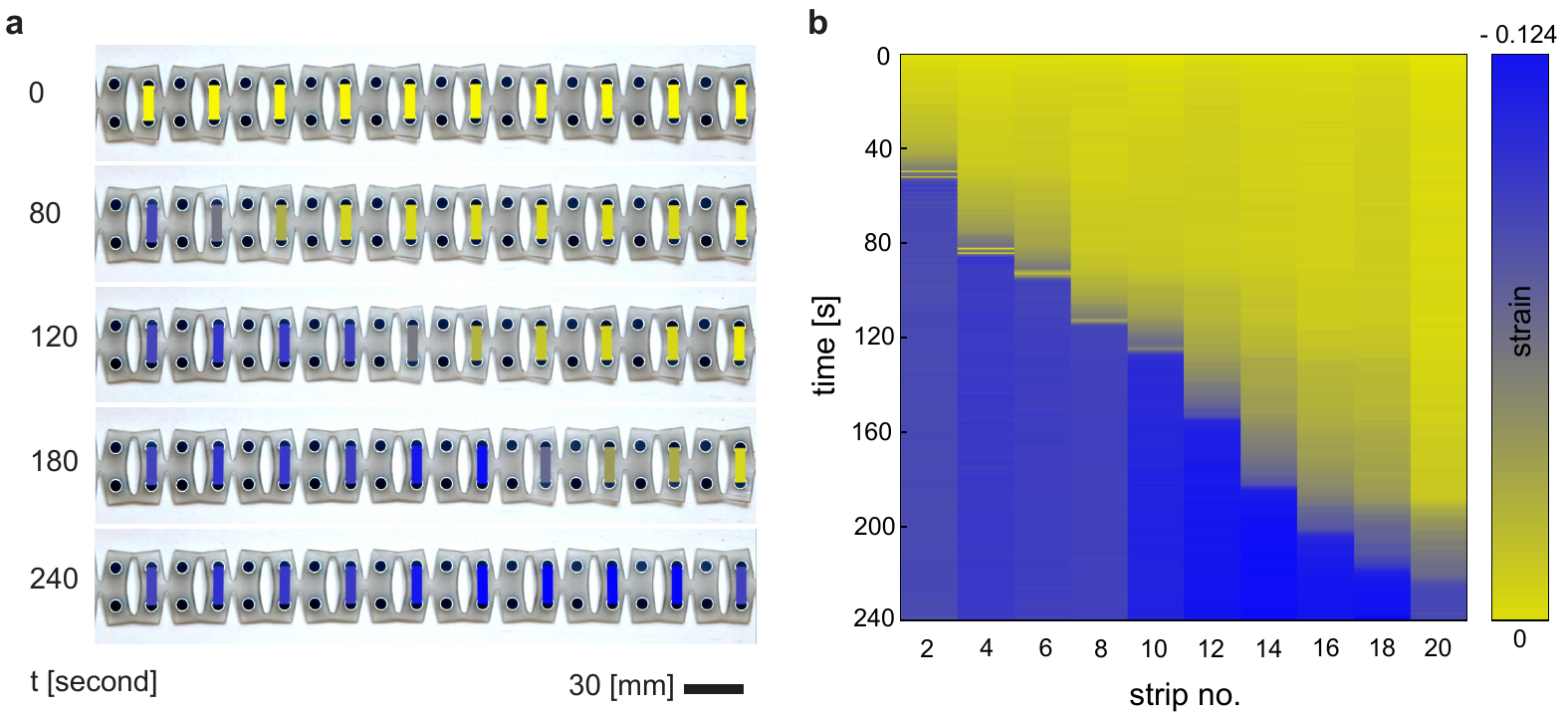}
\caption{{\bf Non-linear travelling kink in a TATA kirigami strip.} (a) The process of snap-back in a kirigami strip. Yellow is corresponding to the high-speed (anti-symmetric) configuration of TATA unit cells and blue represents their low-speed (symmetric) mode. (b) Transition from high-speed mode to low-speed mode}
\label{fig:S5}
\end{figure}

\subsection{Dynamic shape-morphing kirigami}
As an example of functional materials that exhibit dynamic shape-transformation, we showcase a kirigami, made by texturing TAAT unit cells, that transforms its texture over time (\textcolor{blue}{Fig.}~\ref{fig:S6}). 
While computationally we have the freedom to investigate a larger space of material properties, practically, we are limited to available photopolymers (i.e., Agilus and Tango\(^+\)) to fabricate such a kirigami. We, therefore, modulate the strain-rate dependency and the snapping properties of TAAT unit cells by slightly varying their geometry. We found that additional longitudinal short cut-lines (\textcolor{blue}{Fig.}~\ref{fig:S6}) minimizes the stiff interaction of the transversely arranged unit cells. Therefore, the unit cells have more freedom to snap-back while they initially show a high amplitude of out-of-plane buckling---while stretched at high speeds. We, then see that our design (made of 80\% size unit cells) exhibits a dynamic shape-transformation, from a plus- to a minus pattern (\textcolor{blue}{Fig.}~\ref{fig:F4}\textcolor{blue}{f}, \sj{Video V7}) while it is initially stretch at a high speed (1000 \(mm/s\)). 

\begin{figure}[h]
\centering
\includegraphics[width=0.96\textwidth]{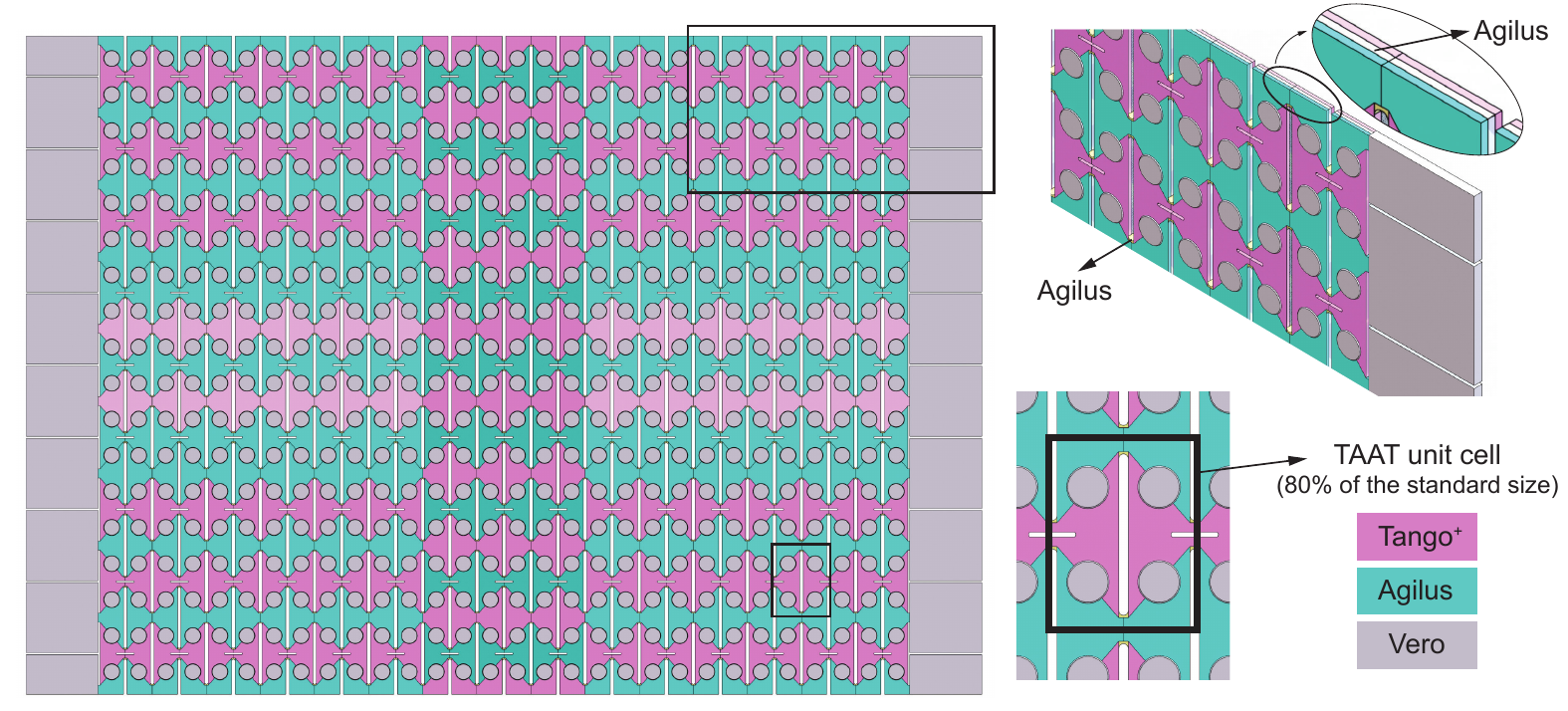}
\caption{{\bf Multi-texture design of the plus-minus shape-morphable plate.} The longitudinal cut-lines minimize the stiff interaction of the transversely arranged unit cells that facilitate the snap-back of TAAT Unit cells}
\label{fig:S6}
\end{figure}

\subsection{Mimicking the behavior of Mimosa pudica}
To mimic the touch-sensitive sequential folding of Mimosa Pudica leaves, we equipped a strip made of 12 identical TATA unit cells (made of standard-size unit cells flipped such that the Agilus finishing layer is at the bottom side) with 3D printed spile connectors that allow taping paper leaflets at the two sides of the snappy half unit cells (\textcolor{blue}{Fig.}~\ref{fig:S7}). The leaflets are light enough so that the kirigami can carry their weight. The two end unit cells do not have leaflets; instead, we glue two foam wedges objects to the snappy half unit cells to prevent the spontaneous wave initiation due to the boundary effects. We then see that TATA unit cells buckle into their anti-symmetric high-speed mode when the strip is stretched at a high speed---that mimics the open configuration of a Mimosa Pudica leaf. Upon touch, the first carrying half unit cell buckles downward, and the following wave emerges mimicking the sequential folding of Mimosa Pudica leaflets (\textcolor{blue}{Fig.}~\ref{fig:F1},
\sj{Video V1}). 
\begin{figure}[h]
\centering
\includegraphics[width=0.96\textwidth]{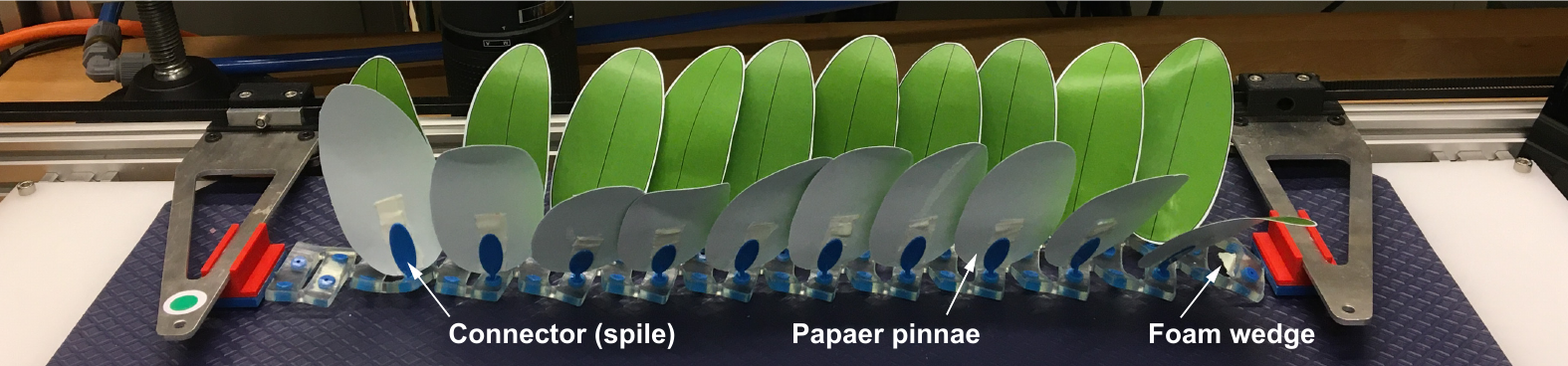}
\caption{{\bf Setup to mimic the behavior of Mimosa pudica.} The assembly of a TATA kirigami strip carrying paper leaflets on our custom-made test setup}
\label{fig:S7}
\end{figure}

\subsection{Mass transportation}
To demonstrate that travelling kinks can move objects, we equipped the 8 middle unit cells of a TATA strip made of 10 unit cells (130\% of standard unit cells) arms and guiders made of rigid plastic (PLA) (\textcolor{blue}{Fig.}~\ref{fig:S8}). In addition, to make sure that the kirigami is able to carry the weight of the plastic arms and of the ping-pong ball while keeping a high enough amplitude of snap-back, we glued a hyperelastic strip on top of the Tango$^{+}$ layer. The hyperelastic strips are mold-cast using an addition silicone rubber (Elite Double 8 Normal, Zhermack) and have been glued to the TATA unit cells using silicone glue (Sil-Poxy, Smooth-On). Now, if we start from a high-speed mode (i.e., anti-symmetric mode), a sequential snapping is able to move a ping-pong ball using PLA arms and guiders(\textcolor{blue}{Fig.}~\ref{fig:F5}, \sj{Video V8}).

\begin{figure}[h]
\centering
\includegraphics[width=0.96\textwidth]{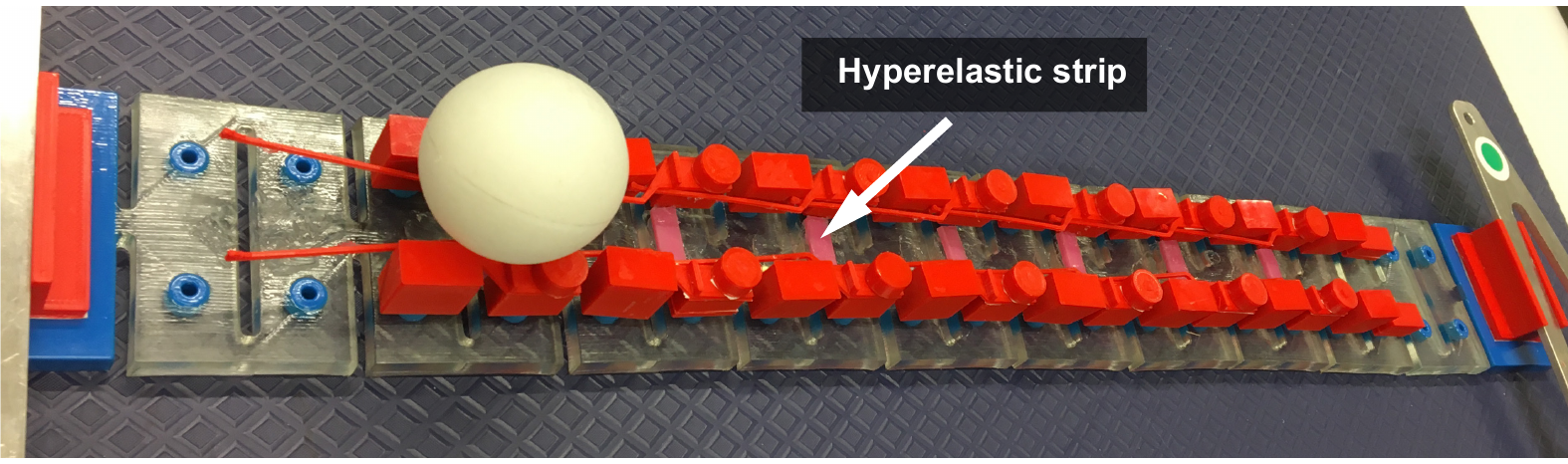}
\caption{{\bf Setup to move a ping-pong ball.} The arms are mounted on the snappy half unit cells which their snap-back amplitude is enhanced using a hyperelastic strip. The non-snappy half unit cells carry the guiders}
\label{fig:S8}
\end{figure}

\subsection{Viscoelastic von Mises trusses}
We use a viscoelastic von Mises truss to model the shape-transformation of individual viscoelastic kirigami unit cells (e.g., from a meta-stable anti-symmetric to a symmetric mode), and wave propagation in kirigami strips.

\subsubsection{\bf Viscoelastic snap-back--0D Model}
The shape-transformation in a kirigami unit cell can be defined by the time-dependent snapping of one of its half-unit cells. Such a change of mode is similar to the pseudo-bistability of a popper geometry \cite{gomez2019dynamics}. Excluding the effect of inertial forces, such time-dependent snapping can be modelled as the snapping of a simple van Mises truss made by joining a pair of linear springs \(K/2\). A pre-stretched spring \(k\) and a dashpot \(c\) with a parallel configuration then connect the moving joint to the ground and transforms the system into a viscoelastic bi-stable configuration (\textcolor{blue}{Fig.}~\ref{fig:S9}). In such a model, the non-linearity is purely geometrical as it comes from the von Mises truss.

\begin{figure}[t!]
\centering
\includegraphics[width=0.98\textwidth]{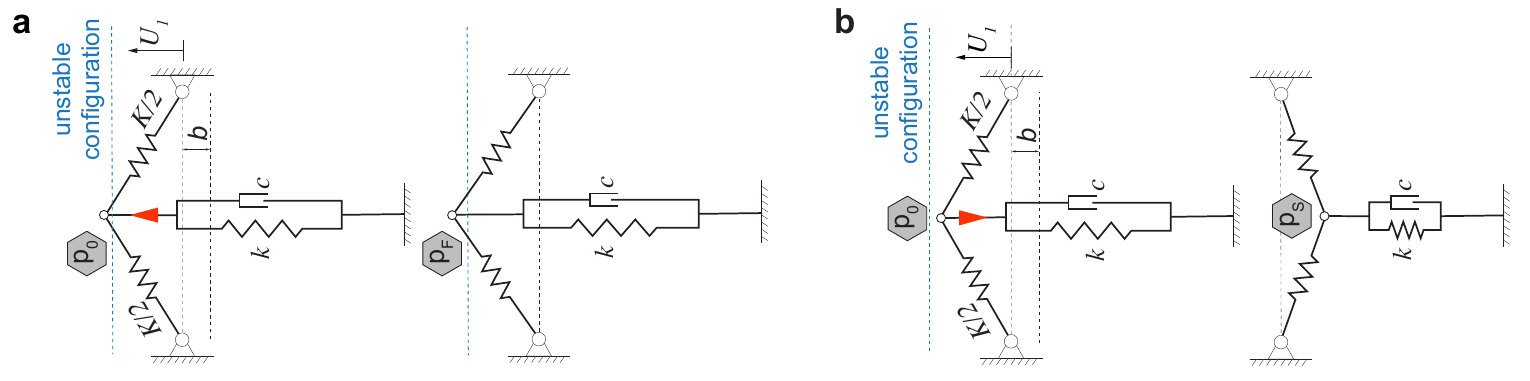}
\caption{{\bf Bi-stable viscoelastic trusses.} The initial condition of a viscoelastic truss determines the (a) movement of the joint in a viscoelastic von Mieses truss towards the high-speed mode or results in a (b) snap-back}
\label{fig:S9}
\end{figure}
To evaluate the dynamics of snapping we used a simplified Lagrangian formulation that comprises the elastic potential energy \(V\) and the Rayleigh dissipation function \(D\):
\begin{equation}
-\pdv{V}{U_i} + \pdv{D}{\dot{U_i}} = 0,
\label{eq:LagEqu}
\end{equation}
\noindent 
where \(U_i\) denotes the displacement of point \(P\) in the generalized coordinate \(i\) (\(i\)=1).
Let's assume that due to an initial condition the joint \(P\) initially moves towards a point (\(P_0\)) close to the unstable configuration. Then, dependent on the offset from the unstable configuration our truss model may exhibit either a movement towards the high-speed stable configuration \(P_F\) (\textcolor{blue}{Fig.}~\ref{fig:S9}\sj{a}) or a snap-back motion towards the low-speed stable configuration \(P_S\) (\textcolor{blue}{Fig.}~\ref{fig:S9}\sj{b}). The potential energy of the truss can be, then, written as:
\begin{equation}
\begin{split}
V(\vec{U}) = \frac{1}{2} {k} (U_1+b)^2 + \frac{1}{2} {K} (\sqrt{{a}^2+{U_1}^2} -d_0)^2,
\end{split}
\label{eq:Potential0D}
\end{equation}
\noindent 
and the Rayleigh dissipation function formulated as:
\begin{equation}
D(\dot{\vec{U}}) = \frac{1}{2} c \dot{U_1}^2
\label{eq:Ryleigh0D}
\end{equation}
\noindent 
where \(d_0\) is the natural length of the springs \(K/2\), \(a\) is half of the distance between the fixed joints and \(b\) is the offset from and straight configuration that describe the free stress configuration of spring \(k\). By substituting the energy terms Eq. (\ref{eq:Potential0D}) and Eq. (\ref{eq:Ryleigh0D}) in Eq. (\ref{eq:LagEqu}), the equations of motion for the viscoelastic snapping can be driven as:
\begin{equation}
c\dot{U_1} = -KU_1 + K  (\frac{d_0}{\sqrt{{a}^2+{U_1}^2}}) U_1 - k (U_1+b)
\label{eq:EqMotion0D}
\end{equation}
\noindent 
Then, to simplify the equation of motion, we use the leading terms of Taylor expansion of nonlinear terms centered at \(X_i=0\) that removes \(U_1\) from the denominator in Eq. (\ref{eq:EqMotion0D}) up to cubic order 3:
\begin{equation}
\dot{U_1} =(-\frac{K} {c}-\frac{k} {c}+\frac{Kd_0} {ca})U_1 - \frac{Kd_0} {2ca^3}U_1^3 - \frac{kb} {c} + \mathcal{O}({U_1}^4)
\end{equation}
To investigate the mechanics of viscoelastic snap-back regardless of geometry and material we nondimensinalize the equation of motion. The dimensionless governing equation of motion can be written as:
\begin{equation}
\bar{U}_{1,\bar{t}} = \bar{U_1} - \beta \bar{U_1}^3 - 1 + \mathcal{O}({\bar{U_1}}^4),
\label{eq:Eq0D-NonDim}
\end{equation}
\noindent
in which $\bar{U}_{1}$ and $\bar{t}$ are the dimensionless displacement and time. The dimensionless parameter $\beta = \frac{Kk^2b^2d_0} {2((-K-k)a+Kd_0)^3}$ is a function of material properties and nonlinear geometry.

\subsubsection{\bf Travelling kink--1D Model}
In the next step, we investigate the mechanics of traveling waves in a kirigami strip which is made by connecting identical 0D trusses using linear elastic springs \(R\) assuring the compliant interaction of moving joints \(P_i\)s (\textcolor{blue}{Fig.}~\ref{fig:S10}).
\begin{figure}[t!]
\centering
\includegraphics[width=0.96\textwidth]{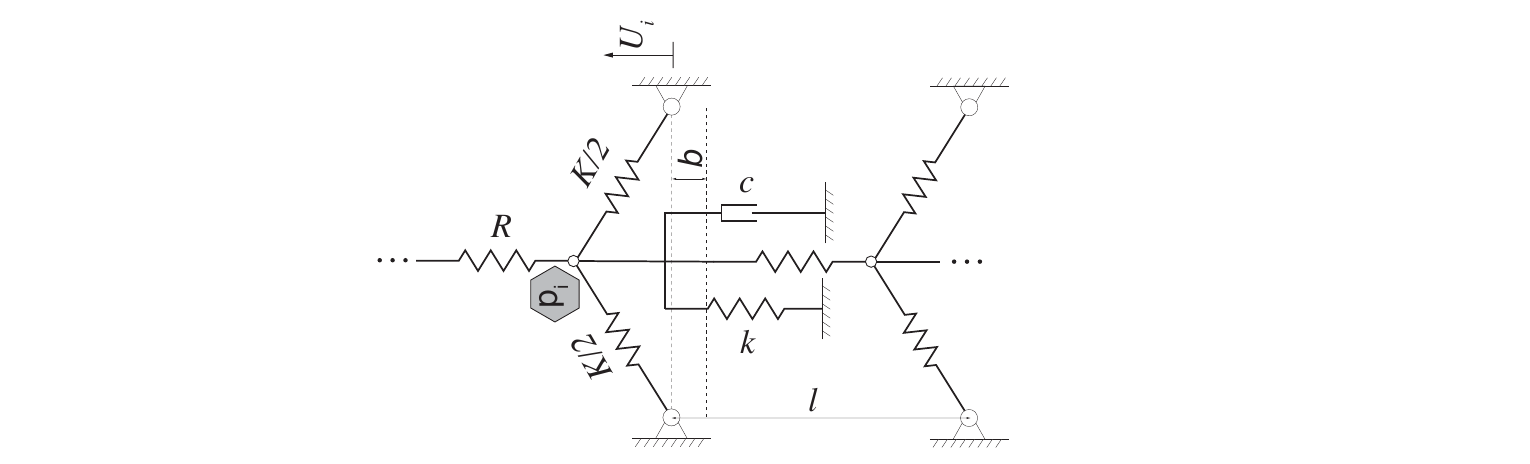}
\caption{{\bf 1D model.} Our 1D model is based on a series of viscoelastic von Mises trusses integrated into a string using the linear springs \(R\)}
\label{fig:S10}
\end{figure}
%
The equation of motion for the \(i\)th joint can be derived in the same way as above, where:
\begin{equation}
\begin{split}
V(\vec{U}) = & \frac{1}{2} {k} \sum_{i=1}^{N} {(U_i+b)}^2 + \frac{1}{2} {K} \sum_{i=1}^{N} (\sqrt{{a}^2+{U_i}^2} -d_0)^2  + \frac{1}{2} {R} \sum_{i=1}^{N-1} ({U_{i+1}}-{U_{i}})^2  \\
& + \frac{1}{2} {R} \sum_{i=1}^{N-1} ({U_{i}}-{U_{i-1}})^2 
\end{split}
\label{eq:Potential1D}
\end{equation}
\noindent 
and the Rayleigh dissipation function is:
\begin{equation}
D(\dot{\vec{U}}) = \frac{1}{2}c \sum_{i=1}^{N} (\dot{U_i})^2
\label{eq:Rayleigh1D}
\end{equation}
\noindent 
The equations of motion for the \(i\)th unit cell can be then written as:
\begin{equation}
c \dot{U_i} = -KU_i + K  (\frac{d_0}{\sqrt{{a}^2+{U_i}^2}}) U_i - k (U_i+b) + R (U_{i+1}-2U_i+U_{i-1})
\label{eq:EqMotion1D}
\end{equation}
\noindent 
Furthermore, in order to simplify the equations of motion, we use the first two terms of Taylor expansion centered at \(X_i=0\) to remove the \(X_i\) from the denominator in Eq. (\ref{eq:EqMotion1D}):
\begin{equation}
\dot{U_i} = (-\frac{K} {c}-\frac{k} {c}+\frac{Kd_0} {ca})U_i - \frac{Kd_0} {2ca^3}U_i^3 - \frac{kb} {c} + \frac{R} {c} (U_{i+1}-2U_i+U_{i-1}) + \mathcal{O}({U_i}^4)
\label{eq:Eq1D-NonDim}
\end{equation}
We then nondimensionalise the equation of motion, whereas by defining \( {U_i} = \frac{kab} {((-K-k)a+Kd_0)} \bar{U}_{i} \) 
the governing equation of motion (14) can be written as:
\begin{equation}
\bar{U}_{i,\bar{t}} = \bar{U}_{i} - \beta \bar{U}_{i}^3 - 1 + L^2 (\bar{U}_{i+1}-2\bar{U}_{i}+\bar{U}_{i-1}) + \mathcal{O}({\bar{U}_i}^4).
\end{equation}
\noindent
in which \(L^2=\frac{Ra}{(-K-k)a+Kd_0}\).
In the continuum limit, the generalized displacements \(\bar{U}_{i+1}\) and \(\bar{U}_{i-1}\) can be rewritten using a Taylor expansion as: ${\bar{U}_{i}}=\bar{U}(x) $ and 
$
{\bar{U}_{i\pm{1}}} =\bar{U}(x) \pm \pdv{\bar{U}}{\bar{x}} + \frac{1}{2} \pdv[2]{\bar{U}}{\bar{x}} \pm \frac{1}{6} \pdv[3]{\bar{U}}{\bar{x}} + \cdots ,
$
\noindent
Therefore, the dimensionless continuum limit of the equation of motion Eq. (\ref{eq:EqMotion1D}) can be written as:
\begin{equation}
\bar{U}_{,\bar{t}} = \bar{U} - \beta {\bar{U}}^3 -1 + \pdv[2]{\bar{U}}{\bar{X}} + \mathcal{O}({U}^4)
\end{equation}
when, \(\bar{X} = L\bar{x}\) is the dimensionless coordinate. 

\end{widetext}
\end{document}